\documentclass[sigconf]{acmart}
\usepackage{subcaption} 
\usepackage[frozencache=true,cachedir=minted-cache-new2]{minted}
\setcopyright{acmcopyright}
\copyrightyear{2021}
\acmYear{2021}
\acmDOI{10.1145/1122445.1122456}

\acmConference[ICAIF'21]{ICAIF'21}{November 03--05, 2021}{London, UK}

\acmISBN{978-1-4503-XXXX-X/18/06}

\begin{document}
\title[ABIDES-Gym]{ABIDES-Gym: Gym Environments for Multi-Agent Discrete Event Simulation and Application to Financial Markets}

\author{Selim Amrouni}
\authornote{Both authors contributed equally to this research.}
\email{selim.amrouni@jpmorgan.com}
\author{Aymeric Moulin}
\authornotemark[1]
\email{aymeric.moulin@jpmorgan.com}
\affiliation{%
  \institution{J.P. Morgan AI Research}
  \streetaddress{383 Madison Avenue}
  \city{New York}
  \state{New York}
  \country{USA}
  \postcode{10179}
}

\author{Jared Vann}
\affiliation{%
  \institution{J.P. Morgan AI Engineering}
  \streetaddress{383 Madison Avenue}
  \city{New York}
  \state{New York}
  \country{USA}
  \postcode{10179}
}
\email{jared.vann@jpmorgan.com}

\author{Svitlana Vyetrenko}
\affiliation{%
  \institution{J.P. Morgan AI Research}
  \streetaddress{383 Madison Avenue}
  \city{New York}
  \state{New York}
  \country{USA}
  \postcode{10179}
}
\email{svitlana.vyetrenko@jpmorgan.com}

\author{Tucker Balch}
\affiliation{%
  \institution{J.P. Morgan AI Research}
  \streetaddress{383 Madison Avenue}
  \city{New York}
  \state{New York}
  \country{USA}
  \postcode{10179}
}
\email{tucker.balch@jpmorgan.com}

\author{Manuela Veloso}
\affiliation{%
  \institution{J.P. Morgan AI Research}
  \streetaddress{383 Madison Avenue}
  \city{New York}
  \state{New York}
  \country{USA}
  \postcode{10179}
}
\email{manuela.veloso@jpmorgan.com}

\renewcommand{\shortauthors}{Amrouni and Moulin, et al.}

\begin{abstract}
Model-free Reinforcement Learning (RL) requires the ability to sample trajectories by taking actions in the original problem environment or a simulated version of it. Breakthroughs in the field of RL have been largely facilitated  by the development of dedicated open source simulators with easy to use frameworks such as OpenAI Gym and its Atari environments. In this paper we propose to use the OpenAI Gym framework on discrete event time based Discrete Event Multi-Agent Simulation (DEMAS). We introduce a general technique to wrap a DEMAS simulator into the Gym framework. We expose the technique in detail and implement it using the simulator ABIDES as a base. We apply this work by specifically using the markets extension of ABIDES, ABIDES-Markets, and develop two benchmark financial markets OpenAI Gym environments for training daily investor and execution agents.\footnote{ABIDES source code is open-sourced on \href{https://github.com/jpmorganchase/abides-jpmc-public}{https://github.com/jpmorganchase/abides-jpmc-public} and available upon request. Please reach out to Selim Amrouni and Aymeric Moulin.} As a result, these two environments describe classic financial problems with a complex interactive market behavior response to the experimental agent's action.
\end{abstract}



\maketitle

\section{Introduction}
    Reinforcement learning (RL) \cite{sutton2018reinforcement} is a field of machine learning that consists in maximizing the objective of an agent. The environment the agent evolves in is modeled by a markov decision process (MDP). The objective is typically defined as a cumulative numerical reward formulation. It is maximized by optimizing the policy used by the agent to choose the actions it takes. 
    
    The world is considered to be divided into two distinct parts: the experimental agent and the rest of the world called the environment. 
    Interactions between the agent and the environment are summarized in: (1) the experimental agent taking actions, (2) the environment evolving to a new state based solely on the previous state and the action taken by the experimental agent.
    At each step the agent receives a numerical reward based on the state and the action taken to reach that state.

    There are two classic types of methods to approach RL problems: model-based methods \cite{pmlr-v28-levine13} and model-free methods \cite{sutton2018reinforcement}. Model-based RL assumes that a model of the state action transition distribution and reward distribution are known. Model-free RL assumes these models are unknown but that instead the agent can interact with the environment and collect samples.
    
    Model-free approaches to RL require the ability for the experimental agent to interact with the MDP environment to gather information about state-action transitions and resulting rewards. This can be done by directly interacting with a real-world system; however, the cost and risk associated with this interaction has been proven to be challenging in most cases. 
    The largest success stories of RL happened with problems where the original target environment is numerical and cheap to run by nature or the environment can be simulated as such (as developed in \cite{DBLP:journals/corr/abs-1904-12901}).   
    
    If an environment is straightforward to model, the reward and new state arising from a state-action pair can be simulated directly. However it is not always the case, there are systems where it is non-trivial to directly model the state and action transition steps. Some of them are by nature multi-agent systems. In that case, the easiest way to model the transition from a state observed by an agent to the next state some time in the future after it took an action is to simulate the actions taken by all the agents in the system.
    
    Discrete Event Multi-Agent Simulation (DEMAS) has been a topic of interest for a long time \cite{ABMDES,simulationhandbook}. There are two main types of DEMAS:
    \begin{itemize}
        \item Time step based simulation: the simulator advances time by increments already determined before starting the simulation, typically of fixed size.
        \item Event time based simulation: the simulator advances time as it processes events from the queue. Time jumps to the next event time.
    \end{itemize}
    In the case of event time based simulation, most of the research has focused on simulating the entire system and its agents and observing the evolution of the different variable of the system.
    
    In this paper, we propose a framework for wrapping an event time based DEMAS simulator into an OpenAI Gym framework. It enables using a multi-agent discrete event simulator as a straightforward environment for RL research. The framework abstracts away the details of the simulator to only present the MDP of the agent of interest. 
    
    To the best of our knowledge, in the context of event time based DEMAS, there has not been any work published where one agent is considered separately with its own MDP and the other agents considered together as the rest of world background that drives the state action transitions and rewards. 
    
    For practical purposes we detail the framework mechanism by applying it to ABIDES \cite{byrd2019abides}
    - a multipurpose multi-agent based discrete event simulator. We illustrate the benefits for RL research by using ABIDES-markets, the markets simulation extension of ABIDES, as a base simulator to build two financial markets trading OpenAI Gym environments and train RL agents in them.

\section{ABIDES: Agent Based Interactive Discrete Event Simulator}
\label{section:ABIDES}
In this section we present the details of the original implementation and use of ABIDES. We introduce the core simulator and its extension to equity markets simulation through ABIDES-Markets.
    \subsection{ABIDES-Core}
    ABIDES is a DEMAS simulator where agents exclusively interact through a messaging system. Agents only have access to their own state and information about the rest of the world from messages they receive. An optional latency model is applied to the messaging system. 

        \subsubsection{Kernel} 
\begin{figure*}[!htb]
            \centering
            \includegraphics[width=.75\textwidth]{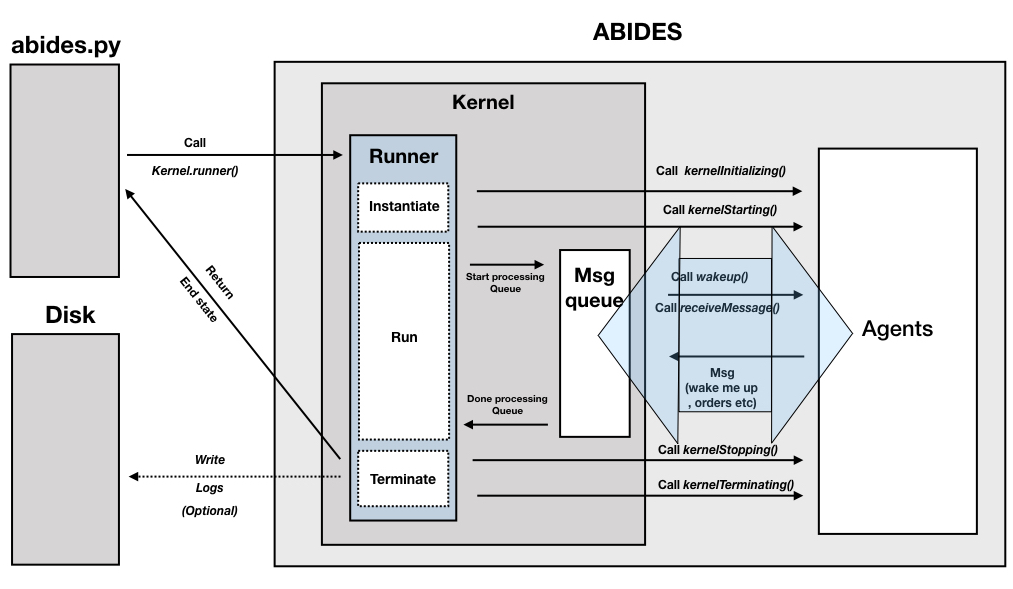}
            \caption{ABIDES-Core kernel mechanism.}
            \label{fig:abides_kernel}
        \end{figure*}

~\\
        \sloppy
        The kernel (see figure \ref{fig:abides_kernel}) drives and coordinates the entire simulation. It is composed of a priority message queue used for handling messages between agents. It takes as input a start time, an end time, a list of agents, a latency model and a pseudo-random seed.
        
        It first sets the clock to the start time, executes the \texttt{kernelInitialize} method for all agents. Then, it calls the \texttt{KernelStarting} method for all agents (this effectively has the same purpose as the \texttt{kernelInitialize} method but has the guarantee that all agents have already been instantiated when it runs).
        
        Agents start sending messages in the initialisation stage. The kernel starts "processing" messages based on reception simulation time (messages are not opened, just routed to the recipient). Messages can be: (1) a general data request message by an agent from another or (2) a wakeup message sent by an agent to itself to be awaken later.
        An agent is only active when it receives a message (general or wakeup). An agent that is active and taking actions will likely result in more messages sent and added to the queue. The kernel will keep processing the message queue until either the queue is empty or simulation time reaches the end time.

        Once the end time has been reached, the kernel calls \texttt{kernelStopping} on all agents then it calls \texttt{kernelTerminating} on all agents. These functions are used to clean and format the data and logs agents have collected throughout the simulation. 
        
        \subsubsection{Agents} ~\\
        An agent is an object with the following methods: \texttt{kernelInitialize}, \texttt{kernelStarting}, \texttt{receiveMessage}, \texttt{kernelStopping}, \texttt{kernelTerminate} and \texttt{wakeUp}. 
        
        Apart from these requirements, agent functioning is flexible. The agent can perform any computations and communicate with the rest of the world by sending messages routed through the kernel.
        
    \subsection{ABIDES-Markets}
        ABIDES-Markets extends ABIDES-Core. It implements a market with a single exchange and several market participants. The exchange is an agent in the simulation and market participants as other agents.

        This way, by construction of ABIDES-core, market participants and the exchange only communicate via messages. Typical messages like orders but also market data rely on the messaging system. (Market data is based on either a direct single request or a recurring subscription)         
        
        This work on ABIDES-Markets has focused on representing the NASDAQ equity exchange and its regular hours continuous trading session \cite{nasdaq_market}. The exchange receives trading instructions similar to the OUCH protocol \cite{nasdaq_ouch}. Orders are matched based on price/time priority model.
        
        The implementation of interactions with the exchange is facilitated by parent object classes FinancialAgent and TradingAgent. The basic background agent inherit from them. They include value agents, momentum agents and others (\cite{byrd2019abides} 
        gives a description of the agents ).

\section{ABIDES-Gym}
In this section we introduce the Gym wrapping. We expose the details of the two-layer wrapping on ABIDES: ABIDES-Gym-Core and ABIDES-Gym sub-environment.
\label{section:ABIDES-Gym}
    \subsection{Motivation}
    As described in section \ref{section:ABIDES}, ABIDES is a flexible tool that facilitates DEMAS. However in its original version, ABIDES presents drawbacks that possibly make it difficult to use for some applications. Creating an experimental agent and adding it to an existing configuration requires a deep understanding of ABIDES. Additionally, the framework is unconventional and makes it hard to leverage popular RL tools. Figure \ref{fig:difference_rl} illustrates the above point: the experimental agent is part of the simulation like the others. For this reason the simulation returns nothing until it is done. The full experimental agent behavior has to be put in the agent code, inside the simulator. There is no direct access to the MDP of the RL problem from outside of ABIDES.

    \subsection{Approach}
    To address the aforementioned difficulties and make ABIDES easily usable for RL we introduce ABIDES-Gym, a novel way to use ABIDES through the OpenAI Gym environment framework. In other words to run ABIDES while leaving the learning algorithm and the MDP formulation outside of the simulator. To the best of our knowledge, it is the first instance of a DEMAS simulator allowing interaction through an openAI Gym framework.
    
\begin{figure*}[!htb]
        \centering
        \includegraphics[width=.75\textwidth]{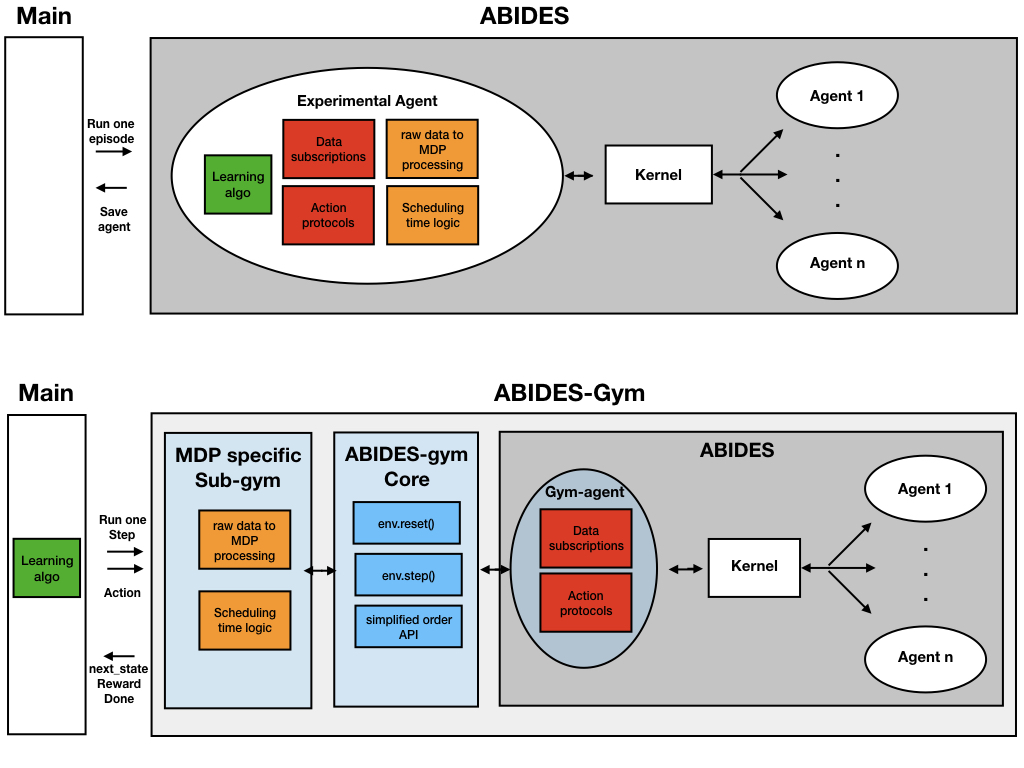}
        \caption{Reinforcement learning framework in ABIDES-Gym vs. regular ABIDES.}
        \label{fig:difference_rl}
    \end{figure*}

    Figure \ref{fig:difference_rl} shows that ABIDES-Gym allows using ABIDES as a black box.
    From the learning algorithm's perspective, the entire interaction with ABIDES-Gym can be summarized into: (1) drawing an initial state by calling $env.reset()$, (2) calling $env.step(a)$ to take an action $a$ and obtain the next state, action, reward and done variables. 
    The sample code listing \ref{listing:code_example_1} shows a short example of the training loop for a learning algorithm.
    
    \begin{listing}[!htb]
    \begin{minted}[fontsize=\footnotesize]{python}
    import gym; import ABIDES_gym
    env = gym.make('markets-daily_investor-v0')
    env.seed(0)
    state, done = env.reset(), False
    agent = MyAgentStrategy(params)
    while not done:
        action = agent.choose_action(state)
        new_state, reward, done, info = env.step(action)
        agent.update_policy(new_state,state,reward,action)
        state = new_state
    \end{minted}
    \caption{Use of Abide-Gym with Open AI Gym APIs}
    \label{listing:code_example_1}
    \end{listing}
    
    \subsection{Key idea: interruptible simulation kernel}
    \label{subsection:key_idea}
    Most DEMAS simulators, including ABIDES, run in one single uninterruptible block. To be able to interact with ABIDES in the Open AI Gym framework we need to be able to start the simulation, pause at specified points in time, return a state and then resume the simulation again.
    
    We propose a new kernel version in which initialization, running and termination phases are broken down into 3 separate methods. The kernel is initialized using the initialization method (effectively calling the \texttt{kernelInitializing} and \texttt{kernelStarting} methods). Then the kernel is run using \texttt{runner} method until either the message queue is empty or an agent sends an interruption instruction to the kernel. When \texttt{runner} finishes, it returns a state from a specified agent. Additionally, we add to the \texttt{runner} method the option to send an action for an agent to execute as first event when simulation resumes. This new kernel can be used in the original mode or in the new Gym mode:
    
    \begin{itemize}
        \item {\bf{Original mode}}. To run ABIDES in the original mode: we successively run the initialization, runner and termination methods. Running a configuration with agents that never send interruptions, the runner method runs until the end of the simulation.
        
        \item {\bf{New Gym mode}}. To run ABIDES in the new Gym mode: we introduce a "placeholder" agent we call Gym agent. At every wake-up call this agent receives, it sends an interruption instruction and its current raw state to the kernel. The kernel pauses the simulation and returns the raw state passed by the Gym agent (the raw state contains all the information passed from the Gym agent to outside of the simulator). The "pause" gives back control to the main user script/thread. The user can use the raw state to perform any computation it wants to select the next action $a$. Using \texttt{runner} with action $a$ as input, the user takes its action and resumes the simulation until the next interruption or the queue is empty. 
    \end{itemize}
    
    \begin{figure*}[!htb]
        \centering
        \includegraphics[width=.65\textwidth]{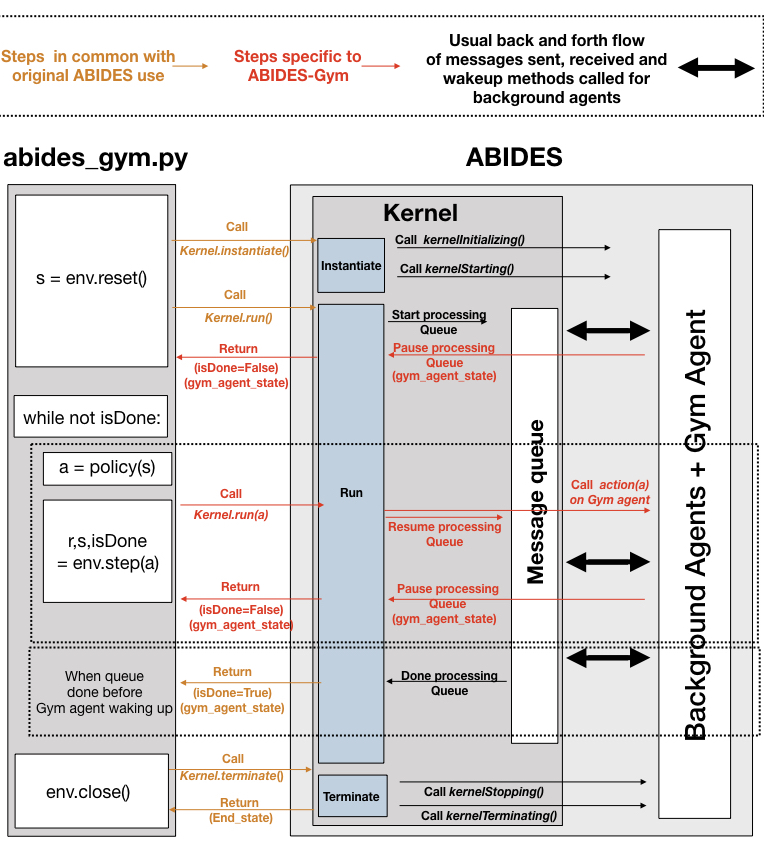}
        \caption{ABIDES-Gym kernel mechanism when running in Gym mode. RL training loop on the left. Describes communications inside between agents an kernel inside ABIDES as well as communications between the RL loop and the simulation. In the figure time is represented by reading the events from top to bottom following the RL training loop on the left}
        \label{fig:abides_gym_kernel}
    \end{figure*}
    
    \subsection{ABIDES-Gym-Core environment: Wrapping ABIDES in OpenAI Gym framework }
    With the new kernel described in the subsection \ref{subsection:key_idea}, ABIDES-Gym wraps ABIDES in an OpenAI Gym framework:
    \begin{itemize}
        \item \texttt{env.reset()}: instantiating the kernel with the configuration, starting simulation using kernel runner method, waiting for the Gym agent to interrupt and send its state, return this state.
        \item \texttt{env.step(a)}: calling runner method on the kernel previously obtained with \texttt{env.reset()} and feeding it action $a$.
    \end{itemize}
    
    This wrapping is independent of the nature of the simulation performed with ABIDES. We structure this into an abstract Gym environment ABIDES-Gym-Core.

    \subsection{ABIDES-Gym sub-environments: Fully defining a Markov Decision Process}
    \label{subsection:defining_MDP}
    ABIDES-Gym-Core abstract environment enforces the gym framework mechanisms. It leaves the MDP undefined. The notions of time steps, state, reward are left undefined. ABIDES-Gym sub-environments, inheriting from ABIDES-Gym-Core, specify these notions as follows:
    
    \begin{itemize}
        \item Time-steps: Gym agent is given a process to follow for its wake-up times (Can be deterministic or stochastic).
        \item State: function is defined to compute actual state of the MDP from the raw state returned by the gym agent.
        \item Reward: function is defined to compute it from the raw state. An additional function is defined to update reward at the end of an episode if needed.
    \end{itemize}
    
    \subsection{More details on ABIDES-Gym-Markets}
    ABIDES-Gym-Markets inherits from ABIDES-Gym-Core and constitutes a middle abstract layer between ABIDES-Gym-Core and environments dedicated to financial markets. In addition to the general wrapping defined by ABIDES-Gym-Core, ABIDES-Gym-Markets handles the market data subscriptions with the exchange, tracking of orders sent, order API and others. In practice we achieve these functionalities by introducing a Gym agent that is more specific to markets simulation.

\section{ABIDES-Gym application to finance: introducing two market environments}
In this section we focus on using the ABIDES-Gym approach for ABIDES-Markets simulations and introduce two market environments to address classic problems. 

    \subsection{Daily Investor Environment}
    This environment presents an example of the classic problem where an investor tries to make money buying and selling a stock throughout a single day. The investor starts the day with cash but no position then repeatedly buy and sell the stock in order to maximize marked to market value at the end of the day (i.e. cash plus holdings valued at the market price).
        \subsubsection{Time steps} ~\\
        \label{subsection:daily_invest_step}
        As described in subsection \ref{subsection:defining_MDP}, we introduce a notion of time step by considering the experimental agent's MDP. Here we make the experimental agent wake up every minute starting at 09:35.
        
        \subsubsection{State space}
        \label{subsection:state_space_daily_invest}
        ~\\
        The experimental agent perceives the market through the state representation:
        $$s(t)=(holdings_t, imbalance_t, spread_t, directionFeature_t, R^k_t) $$
        where:
        \begin{itemize}
            \item $holdings_t$: number of shares of the stock held by the experiment agent at time step $t$
            \item $imbalance_t=\frac{bids \ volume}{ bids \ volume + asks \ volume}$ using the first 3 levels of the order book. Value is respectively set to $0, 1, 0.5$ for no bids, no asks and empty book.
            \item $spread_t=bestAsk_t-bestBid_t$
            \item $directionFeature_t= midPrice_t-lastTransactionPrice_t$
            \item $R^k_t=(r_t,...,r_{t-k+1})$ series of mid price differences, where $r_{t-i}=mid_{t-i}-mid_{t-i-1}$. It is set to 0 when undefined. By default $k=3$ 
        \end{itemize}
        
        \subsubsection{Action space} ~\\
        The environment allows for 3 simple actions: "BUY","HOLD" and "SELL". The "BUY" and "SELL" actions correspond to market orders of a constant size, which is defined at the instantiation of the environment. It is is defaulted to 100.
        
        \subsubsection{Reward} ~\\
        We define the step reward as:
        $reward_t=markedToMarket_t-marketToMarket_{t-1}$ where $marketToMarket_t=cash_t+holdings_t\cdot lastTransaction_t$ and $lastTransaction_t$ the price at which the last transaction in the market was executed before time step $t$.
        
    \subsection{Algorithmic Execution Environment}
    This environment presents an example of the algorithmic order execution problem. The agent has either an initial inventory of the stocks it tries to trade out of or no initial inventory and tries to acquire a target number of shares. The goal is to realize this task while minimizing transaction cost from spreads and market impact. It does so by splitting the parent order into several smaller child orders. In \cite{kearns_execution}, the problem of designing optimal execution strategies using RL in an environment with static historical market data was considered.
    
    \subsubsection{Definitions} ~\\
    The environment has the following parameters and variables:
    \begin{itemize}
        \item $parentOrderSize$: 
        Total size the agent has to execute (either buy or sell). It is defaulted to 20000.
        \item $direction$: direction of the $parentOrder$ (buy or sell). It is defaulted to buy.
        \item $timeWindow$: Time length the agent is given to proceed with $parentOrderSize$ execution. It is defaulted to 4 hours.
        \item $childOrderSize$ the size of the buy or sell orders the agent places in the market. It is defaulted to 50.
        \item $startingTime$ time of the first action step for the agent
        \item $entryPrice$ is the $midPrice_t$ for $t=startingTime$
        \item 
        $nearTouch_t$ is the highest $bidPrice$ if $direction = buy$ else is the lowest $askPrice$
        \item 
        $penalty$: it is a constant penalty per non-executed share at the end of the $timeWindow$. It is defaulted to 100 per share.
    \end{itemize}

        \subsubsection{Time steps} ~\\
        We use the same notion of time steps as described in \ref{subsection:daily_invest_step}. Here the agent wakes up every ten seconds starting at 09:35 ($startingTime$).
        
        \subsubsection{State Space} ~\\
        The experimental agent perceives the market through the state representation:
        \begin{align*}
        s(t)=(& holdingsPct_t, timePct_t,\\ &differencePct_t, imbalance5_t, imbalanceAll_t,\\ &priceImpact_t, spread_t, directionFeature_t, R^k_t)
        \end{align*}
        where:
        \begin{itemize}
            \item $holdingsPct_t = \frac{holdings_t}{parentOrderSize}$: the execution advancement
            \item
            $timePct_t=\frac{t - startingTime}{timeWindow}$: the time advancement
            \item 
            $differencePct_t = holdingsPct_t - timePct_t$ 
            \item 
            $priceImpact_t = midPrice_t - entryPrice$
            \item
            $imbalance5_t$ and $imbalanceAll_t$ are defined such as in \ref{subsection:state_space_daily_invest} but taking the first 5 levels and all levels of the Order Book.
            \item $spread_t$, $directionFeature_t$, $R^k_t$ defined in  \ref{subsection:state_space_daily_invest} with $k=3$ 
        \end{itemize}
        
        \subsubsection{Action space} ~\\
        \sloppy
        The environment allows for three simple actions: "MARKET ORDER", "DO NOTHING" and "LIMIT ORDER". They are defined as follow:
        \begin{itemize}
            \item "MARKET ORDER": the agent places a market order of size $childOrderSize$ in the direction $direction$. (Instruction to buy or sell immediately at the current best available price)
            \item "LIMIT ORDER": the agent places a limit order of size $childOrderSize$ in the direction $direction$ at the price level $nearTouch_t$. (Buy or sell only at a specified price or better, does not guarantee execution)
            \item "DO NOTHING": no action is taken.
            \end{itemize}
            Before sending a "MARKET ORDER" or a "LIMIT ORDER", the agent will cancel any living order still in the Order Book.
        
    \subsubsection{Reward} ~\\
    
    We define the step reward as:
        $reward_t= \frac{PNL_t}{parentOrderSize}$\\
    with:
    \begin{align*}
    PNL_t = \sum_{o \in O_t } numside\cdot(entryPrice - fillPrice_{o})* quantity_{o} \\
    \end{align*}
    where $numside=1$ if direction is buy else $numside=0$ and $O_t$ is the set of orders executed between step $t-1$ and $t$
    
    We also define an episode update reward that is computed at the end of the episode. Denoting $O^{episode}$ the set of all orders executed in the episode, it is defined as: 
    \begin{itemize}
        \item 0 if $\sum_{o \in O^{episode}} quantity_{o} = parentOrderSize$
        
        \item else $|penalty \times ( parentOrderSize - \sum_{o \in O^{episode}} quantity_{o})|$
        
    \end{itemize}

\section{Experimental Example: Training a reinforcement learning agent in our environments }

To illustrate the ease of use and that agents are able to learn in the environment, we train a learning agent using the suite of tools built on the top of Ray \cite{moritz2018ray}.

RL training loops for Gym environments are very similar and there are widely used standard models. Tune \cite{liaw2018tune} enables us to only input our environment, its parameters and the name of the standard RL algorithm (implemented in RLlib \cite{pmlr-v80-liang18b}) we want to train on it. The Listing \ref{listing:code_example_2} illustrates an example
for training a Deep Q-Learning (DQN)\cite{mnih2013atari} algorithm. 

    \begin{listing}
    \begin{minted}[fontsize=\footnotesize]{python}
    from ABIDES_gym.envs.markets_daily_investor_environment_v0 \
    import SubGymMarketsDailyInvestorEnv_v0
    import ray; from ray import tune
    ray.init()
    tune.run(
        "DQN",
        name="dqn_training",
        stop={ "training_iteration": 200}, #train 200k steps
        checkpoint_freq = 40, #snapshot model every 40k steps
        config={
            # Environment Specification
            "env": SubGymMarketsDailyInvestorEnv_v0, #env used
            "env_config": {'ORDER_FIXED_SIZE':100, 
                            # 1min wakeup frequency
                          'TIMESTEP_DURATION':{'seconds':1*60}},
            "seed": tune.grid_search([1,2,3]), #3 seeds
            # Learning algorithm specification
            "hiddens":[50, 20],
            "gamma": 1, #no discounting},
    )
    \end{minted}
    \caption{Use of Tune APIs}
    \label{listing:code_example_2}
    \end{listing}

    \subsection{Daily Investor environment}
        \subsubsection{Setup}
        \label{subsubsection:setup_daily_vestor}
        \begin{itemize}
            \item Objective function: non-discounted sum of rewards
            \item Algorithm: Deep Q-Learning.
            \item Architecture: 1 fully connected feed forward neural network with 2 layers composed of 50 and 20 neurons.
            \item Learning rate schedule: linear decrease from $1 \cdot 10^{-3}$ to 0 in 90k steps.
            \item Exploration rate: $\epsilon$-Greedy search from 1 to 0.02 in 10k steps.
        \end{itemize}
    
        \subsubsection{Results}
\begin{figure}[!b]
        \centering
        \includegraphics[width=.5\textwidth]{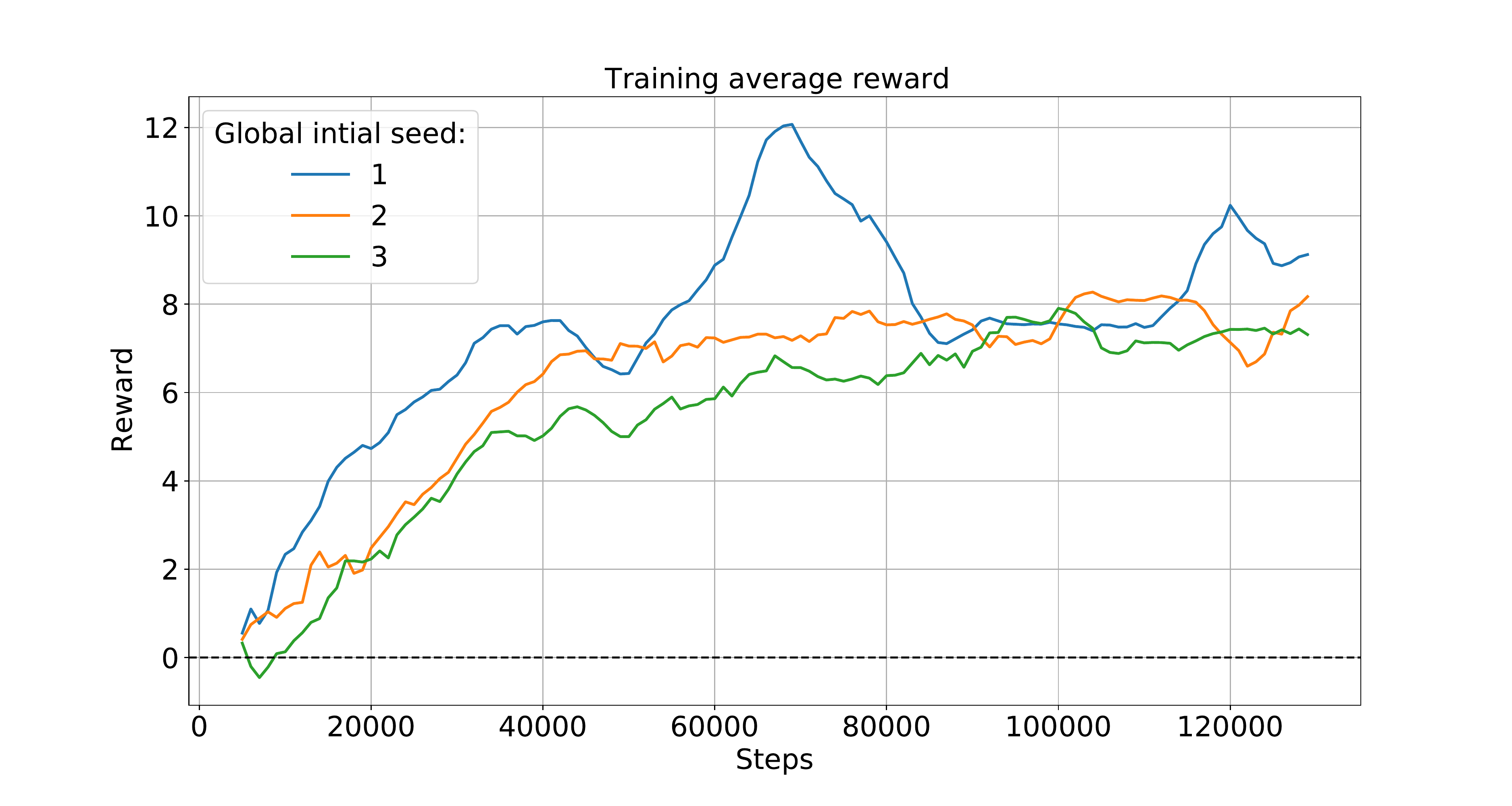}
        \caption{Daily Investor env: training of a DQN agent (average reward per 1000 with 5-point moving average)}
        \label{fig:training_reward}
        \end{figure}

~\\
        Figure \ref{fig:training_reward} shows the average reward throughout training for different initial environment seeds (1, 2 and 3). For all of these different global initial seeds the agent is able to learn a profitable strategy.
        
    \subsection{Execution environment}
        \subsubsection{Setup}~\\
        Same setup as in \ref{subsubsection:setup_daily_vestor} except for the learning rate fixed at $1 \cdot10^{-4}$.
    
        \subsubsection{Results}
\begin{figure}[!t]
        \centering
        \includegraphics[width=.5\textwidth]{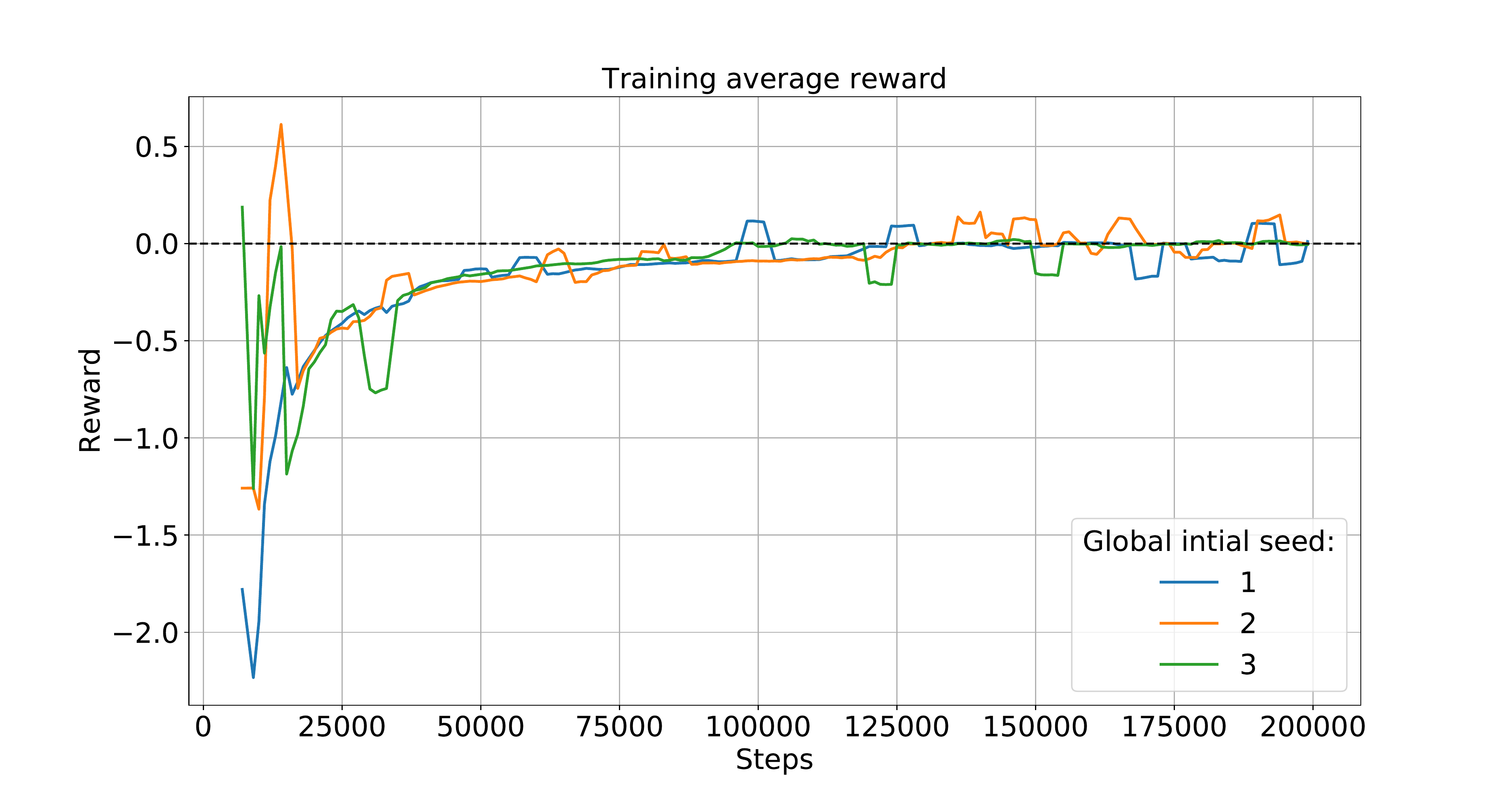}
        \caption{Daily Execution env: training of a DQN agent (average reward per 1000 with 5-point moving average)}
        \label{fig:training_reward_exec}
        \end{figure}

~\\
        Figure \ref{fig:training_reward_exec} shows the average reward throughout training for different initial environment seeds (1, 2 and 3). For the execution problem, achieving a reward close to 0 means executing at a low cost. The agent indeed learns to execute the parent order while minimizing the cost of execution. At the moment there are positive spikes explained by the rare unrealistic behaviors of the synthetic market.

\section{Related Work}
    We now briefly discuss related work on simulation approaches, Reinforcement Learning, OpenAI Gym Environment and applications to trading environment. 
    \subsection{DEMAS simulators}
    DEMAS enables modeling the behavior and evolution of complex systems. It is sometimes easier to generate samples or estimate statistics from a distribution by reproducing the components' behavior rather than trying to directly express the distribution or directly produce a sample. 
    Multiple open-source simulators exist, however none of them has become a de-facto standard. Many of them contain specifics of the domain they arose from. For example, the Swarm simulator \cite{Minar96theswarm} is well suited to describe biological systems. It has the notion of swarm, group of agents and their schedule, which can in turn be connected to build a configuration. For this paper we choose to use ABIDES \cite{byrd2019abides} 
    as a base because of its message based kernel that enables precise control of information dissemination between agents. This makes ABIDES particularly well suited for the simulation of financial markets. 
    
    Even though the above described simulators enable modeling complex systems they are not specifically designed for RL. Many different MDPs can be formulated depending on the part of the whole system chosen as experimental agent and the considered task and reward formulation. Thus, they often lack the ability to interact with the simulator via an easy API to train one or more experimental RL agents with the rest of the system as a MDP background.
    
    \subsection{Multi-agent RL}
    \label{subsection:MARL}
    As we deal with RL in multi-agent scenarios it is important to mention Multi-Agent RL (MARL).
    MARL is a sub-field of RL where several learning agents are considered at once interacting with each other.
    On the one hand MARL could be considered as a more general instance of the problem we introduce in this paper as all agents (or at least most) are learning agents in MARL. On the other hand, so far the problem formulations that have been used for MARL can be quite restrictive for the nature of agent interaction as described below.
    Classic MARL problem formulations used: 
    \begin{itemize}
        \item Partially Observable Stochastic Games (POSGs) \cite{Littman94markovgames}: This formulation is widely used. All agents "take steps" simultaneously by taking an action and observing the next state and reward. It is inherently time-step based -- one needs to consider all time-steps and provide many null actions if the game is not fully simultaneous by nature. Additionally, environments modeling these type of games are typically designed to have all the players be learning agents and have no notion of a "rest of the world" background. 
        
        \item Extensive Form Games (EFGs) \cite{EFGs,RePEc:eee:gamhes:1}: 
        These games are represented by a decision tree and are typically used to model inherently sequential games like chess. Agents play sequentially and a "nature" agent can be added in order to model randomness in the game. This modelling is flexible but still typically imposes constraints on agents' turns to play. E.g. 2 indiscernible nodes of the game from observation should have the player play after them. \cite{LanctotEtAl2019OpenSpiel} Introduces OpenSpiel a tool to create and use existing game implementations, including EFGs. While it seems like one of the most flexible open source tools for multi-agent RL, it is geared towards simpler theoretical games and not adapted to the purpose developed in our work. 
 
    \end{itemize}
    Overall MARL is concerned with learning optimal policies for a group of agents interacting with each other in a game, potentially including a nature agent. 
    
   \subsection{OpenAI Gym environments}
    OpenAI Gym \cite{brockman2016openai} introduced one of the most widely used frameworks and family of environments for RL.
    Its success can be explained by the simplicity of use of Gym environments and how clearly they define the MDP to solve.
    
    Among others, many environments based on classic Atari arcade games have been developed, open sourced and constitute reference benchmarks for the field. 
        \subsubsection{Single agent} ~\\
        Most of the environments are single agent. The problem consists in controlling the actions of a single agent evolving in the environment and maximize its rewards. E.g. CartPole-v1, need to keep a pole standing by controlling its base.
        
        \subsubsection{Multi-agent} ~\\
        Some "multi-agent" environments are provided. They are MARL environments as described in \ref{subsection:MARL}. They allow controlling several learning agents interacting with each other. E.g. PongDuel-v0 where 2 agents play Atari style pong.
    
    OpenAI and third parties provide ready to use environments, the Gym concept provides a framework for how to present the environment to the user. However, Gym does not provide constraints on how to design the algorithm producing the environment transitions.

    \subsection{OpenAI Gym trading environments }
    RL environments for trading are a good illustration of the transition modelling issue. The only registered OpenAI Gym trading environment is \cite{gym-anytrading}. It provides the typical Gym ease of use but the transitions of the MDP are generated by replaying historical market data snapshots. While replaying historical market data to assess an investment strategy is classic, it is done under the assumption that trades entered by the experimental agent does not impact future prices. This assumption would make the use of RL unnecessary since the experimental agent's actions have no impact on the environment. 
    
    \subsection{Financial markets dynamics modeling using multi-agent simulation}
    DEMAS has been used to model markets dynamics. Earlier work like \cite{nature-multiagent} focused on reproducing time scaling laws for returns.
    In recent contributions, 
    \cite{byrd2019abides,vyetrenko2019real} 
    study the impact of a perturbation agent placing extra orders into a multi-agent background simulation and compare the differences in observed prices. In \cite{vyetrenko2019real,balch2019evaluate}
    the benefits of DEMAS market simulation against market replay are developed by proposing experimental evidence using ABIDES.

\begin{figure}[!b]
    \centering
    \includegraphics[width=.45\textwidth]{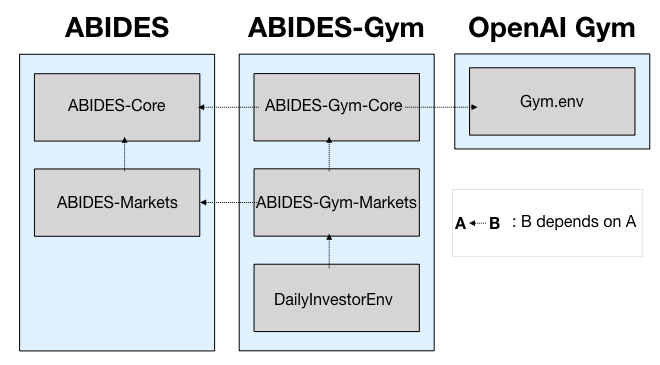}
    \caption{Dependency Diagram}
    \label{fig:dependency_diagram}
\end{figure}

\section*{Conclusion}
    Our contributions are three fold: (1) Provide a general framework to wrap a DEMAS in a Gym environment. (2) Develop the framework in details and implementing it on ABIDES simulator.  (3) Introduce two financial markets environments: DailyInvestorEnv and ExecutionEnv as benchmarks for supporting research on interactive financial markets RL problems.
    Figure \ref{fig:dependency_diagram} illustrates the dependencies between ABIDES, ABIDES-Gym and Open AI Gym.
    
In this work, we introduced a technique to wrap a DEMAS simulator into the OpenAI Gym environment framework. We explicitly used this technique on the multi-purpose DEMAS simulator ABIDES to create ABIDES-Gym. We used ABIDES-Gym and ABIDES's markets extension ABIDES-Markets to build the more specific ABIDES-Gym-Markets, an abstract Gym environment constituting a base for creating financial markets Gym environment based on ABIDES-Markets. Based on it we introduced two new environments for training an investor and an execution agent. In addition, by leveraging open-source RL tools, we demonstrated that an RL agent could easily be trained using ABIDES-Gym-Markets.\\


\section{Acknowledgments}

We would like to thank Yousef El-Laham and Vineeth Ravi for there contributions.\\

Disclaimer:

This paper was prepared for informational purposes by the Artificial Intelligence Research group of JPMorgan Chase \& Co\. and its affiliates (``JP Morgan''), and is not a product of the Research Department of JP Morgan. JP Morgan makes no representation and warranty whatsoever and disclaims all liability, for the completeness, accuracy or reliability of the information contained herein. This document is not intended as investment research or investment advice, or a recommendation, offer or solicitation for the purchase or sale of any security, financial instrument, financial product or service, or to be used in any way for evaluating the merits of participating in any transaction, and shall not constitute a solicitation under any jurisdiction or to any person, if such solicitation under such jurisdiction or to such person would be unlawful.

\normalsize


\bibliographystyle{ACM-Reference-Format}
\bibliography{main}

\end{document}